# From laboratory experiments to LISA Pathfinder: achieving LISA geodesic motion.

F Antonucci[a], M Armano[b], H Audley[c], G Auger[d], M Benedetti[e], P Binetruy[d], C Boatella[f], J Bogenstahl[c], D Bortoluzzi[g], P Bosetti[g], N Brandt[h], M Caleno[i], A Cavalleri[a], M Cesa[i], M Chmeissani[j], G Ciani[k], A Conchillo[l], G Congedo[a], I Cristofolini[g], M Cruise[m], K Danzmann[c], F De Marchi[a], M Diaz-Aguilo[n], I Diepholz[c], G Dixon[m], R Dolesi[a], N Dunbar[y], J Fauste[b], L Ferraioli[a], D Fertin[i], W Fichter[o], E Fitzsimons[p], M Freschi[b], A García Marin[c], C García Marirrodriga[i], R Gerndt[h], L Gesa[l], D Giardini[q], F Gibert[l], C Grimani[r], A Grynagier[o], B Guillaume[i], F Guzmán[s], I Harrison[t], G Heinzel[c], M Hewitson[c], D Hollington[u], J Hough[p], D Hoyland[m], M Hueller[a], J Huesler[i], O Jeannin[d], O Jennrich[i], P Jetzer[v], B Johlander[i], C Killow[p], X Llamas[w], I Lloro[l], A Lobo[l], R Maarschalkerweerd[t], S Madden[i], D Mance[v], I Mateos[l], P W McNamara[i], J Mendes[t], E Mitchell[u], A Monsky[c], D Nicolini[i], D Nicolodi[a], M Nofrarias[c], F Pedersen[i], M Perreur-Lloyd[p], A Perreca[a], E Plagnol[d], P Prat[d], G D Racca[i], B Rais[d], J Ramos-Castro[x], J Reiche[c], J A Romera Perez[i], D Robertson[p], H Rozemeijer[i], J Sanjuan[k], A Schleicher[h], M Schulte[u], D Shaul[u], L Stagnaro[i], S Strandmoe[i], F Steier[c], T J Sumner[u], A Taylor[p], D Texier[b], C Trenkel[y], D Tombolato[a], S Vitale[a], G Wanner[c], H Ward[p], S Waschke[u], P Wass[u], W J Weber[a], P Zweifel[q].

[a] Dipartimento di Fisica, Università di Trento and INFN, Gruppo Collegato di Trento, 38050 Povo, Trento, Italy
[b] European Space Astronomy Centre, European Space Agency, Villanueva de la Ca˜nada, 28692 Madrid, Spain
[c] Albert-Einstein-Institut, Max-Planck-Institut für Gravitationsphysik und Universität Hannover, 30167 Hannover, Germany
[d] APC UMR7164, Université Paris Diderot, Paris, France.
[e] Dipartimento di Ingegneria dei Materiali e Tecnologie Industriali, Università di Trento and INFN, Gruppo Collegato di Trento, Mesiano, Trento, Italy
[f] CNES, DCT/AQ/EC, 18 Avenue Edouard Belin, 31401 Toulouse, Cédex9, France
[g] Dipartimento di Ingegneria Meccanica e Strutturale, Università di Trento and INFN, Gruppo Collegato di Trento, Mesiano, Trento, Italy
[h] Astrium GmbH Claude-Dornier-Straße 88090 Immenstaad Germany
[i] European Space Technology Centre, European Space Agency, Keplerlaan 1, 2200 AG Noordwijk, The Netherland
[j] IFAE, Universitat Autònoma de Barcelona, E-08193 Bellaterra (Barcelona), Spain
[k] Department of Physics, University of Florida, Gainesville, FL 32611-8440, USA
[l] ICE-CSIC/IEEC, Facultat de Ciències, E-08193 Bellaterra (Barcelona), Spain
[m] Department of Physics and Astronomy, University of Birmingham, Birmingham, UK
[n] UPC/IEEC, EPSC, Esteve Terrades 5, E-08860 Castelldefels, Barcelona, Spain
[o] Institut für Flugmechanik und Flugregelung, 70569 Stuttgart, Germany
[p] Department of Physics and Astronomy, University of Glasgow, Glasgow, UK
[q] Institut für Geophysik, ETH Zürich, Sonneggstrasse 5, CH-8092, Zürich, Switzerland
[r] Istituto di Fisica, Università degli Studi di Urbino/ INFN Urbino (PU), Italy
[s] NASA -Goddard Space Flight Centre, Greenbelt, MD 20771, USA
[t] European Space Operations Centre, European Space Agency, 64293 Darmstadt, Germany
[u] The Blackett Laboratory, Imperial College London, UK
[v] Institut für Theoretische Physik, Universität Zürich, Winterthurerstrasse 190, CH-8057 Zürich, Switzerland
[w] NTE-SENER, Can Malé, E-08186, Lli¸cà d'Amunt, Barcelona, Spain

x Universitat Politècnica de Catalunya, Enginyeria Electrònica, Jordi Girona 1-3, 08034 Barcelona, Spain
y Astrium Ltd, Gunnels Wood Road, Stevenage, Hertfordshire, SG1 2AS, UK

E-mail: Stefano.Vitale@unitn.it

**Abstract.** This paper presents a quantitative assessment of the performance of the upcoming LISA Pathfinder geodesic explorer mission. The findings are based on the results of extensive ground testing and simulation campaigns using flight hardware and flight control and operations algorithms. The results show that, for the central experiment of measuring the stray differential acceleration between the LISA test masses, LISA Pathfinder will be able to verify the overall acceleration noise to within a factor two of the LISA requirement at 1 mHz and within a factor 6 at 0.1 mHz. We also discuss the key elements of the physical model of disturbances, coming from LISA Pathfinder and ground measurement, that will guarantee the LISA performance.

# 1 Introduction

LISA Pathfinder (LPF) is a precursor mission to LISA, with the scope of demonstrating that the hardware designed for LISA can achieve both geodesic motion, and interferometric tracking of free-falling test-masses (TMs), at the level of purity required by LISA scientific requirements. The mission, the details of its instruments, and its relation to LISA have been described in various papers [1]. Its development status is described by an accompanying paper in this same issue of the journal [4].

The mission consists of a series of experiments aimed at measuring a set of physical quantities that underpin the LISA performance budget. The most important of these experiments is the measurement of the Power Spectral Density (PSD) of parasitic forces that cause a differential acceleration noise between the two free-falling TMs at the end points of each LISA arm. This acceleration competes with gravitational wave signals at the lower end of LISA sensitivity band.

Many other experiments are dedicated to quantitatively identifying the physical sources of force disturbance, with the goal of achieving a full projection of the differential acceleration noise into its components, with. a quantitative estimate of the PSD as a sum of all leading, independent contributions. The projection is satisfactory if the residual mismatch between the sum of these estimated contributions, and the measured PSD of acceleration noise is less than the measurement errors. These experiments have been designed in detail over the last few years, and their expected performance has been estimated analytically [5] in the past. They are now in the course of being simulated within the mission end-to-end simulator [7], allowing verification of those initial predictions.

Furthermore, in parallel with the experiment simulation, flight models, or at least fully representative qualification models, have been delivered for all of the hardware that may affect the mission performance. This has allowed an intense testing campaign in the laboratory. The results of this campaign have a twofold impact: on one side they validate the estimate of the in-flight performance of LPF. On the other, they give a direct measurement of key parameters of the physical model for some of the expected disturbances. This allows for a direct extrapolation to LISA, in many cases also for frequencies below 1 mHz, the lower end of LPF measurement band.

The paper reports on the results of these simulation and testing campaigns, and discusses their consequences in estimating the expected performance of LPF. It also discusses how, if the performance is indeed achieved, the results from LPF and ground testing may be combined to estimate LISA performance.

# 2 LISA Pathfinder principles.

As stated above, the details of LPF instrument and mission can be found in [1] [2], [3]. A summary of the basic concepts is the following.

- The core of LPF is a down-scaled version of one LISA arm, called the LISA Technology Package (LTP) consisting of two TMs having no mechanical contact to the SC they are both enclosed in, and thus being in nominal free fall.

- The relative displacement of TMs and SC along a single shared axis, that we call x, are measured by two, pm-level accuracy, laser interferometers. More precisely, one interferometer measures the displacement $x_1$ of one of the TMs relative to SC, and the second interferometer measures the relative displacements $x_{12}$ between the TMs.
- Two control loops force, respectively, the SC and one of the two TMs (TM2), to maintain fixed distances to the remaining TM (TM1), which thus serves as a geodesic reference. The first control loop is called the drag-free (DF) loop and acts via a set of micro-thrusters that apply forces to the SC. The second is called the Electrostatic Suspension (ES) loop and acts via the electrodes that surround TM2.
- Most of the hardware used on LPF is identical to that for LISA [1]. In particular the TMs and their surrounding apparatus, called the Gravity Reference Sensor (GRS), including electrodes, electrode housing, TM launch lock and release mechanism, UV-light discharging system, Tungsten masses for gravitational balance, and vacuum enclosure, are equal to those for LISA. The laser interferometer that measures the position of TM1 relative to the SC, though of different design, has the same performance required for that to be used in LISA for the same purpose. Finally the micro-thrusters used to control the SC have same performance as those of LISA, though they have only been qualified for the shorter lifetime of LPF.

The objectives of the mission are, in essence, those of:

- Showing that parasitic forces are sufficiently small, such that the relative acceleration of the two TMs will have a PSD$^{1/2}$ less than $3\times10^{-14}\,\mathrm{m\,s^{-2}}/\sqrt{\mathrm{Hz}}$ at 1 mHz. This figure is larger than LISA requirement at the same frequency $\sqrt{2}\times3\times10^{-15}\,\mathrm{m\,s^{-2}}/\sqrt{\mathrm{Hz}}$, by a factor 7. In addition LISA must maintain this requirement down to 0.1 mHz.
- Accounting for the measured acceleration PSD. To be specific, the measured PSD must be apportioned to the contributions due to the various expected physical sources, each contribution having been quantitatively demonstrated by tests performed either on board or on ground. The final uncertainty in accounting for the observed noise, gives the residual uncertainty on the PSD of parasitic forces acting on LISA TM, that have not been modeled within our current understanding of the apparatus.
- Showing that the relative motion of the centers of mass of two free-falling TMs, and that of each of these centers of mass relative to a SC fixed frame, can be tracked along a common direction, with an accuracy of better than 10 pm/√Hz, at frequencies between 3 mHz and 1 Hz. This is the same as the LISA requirement, though in LISA the TM-to-TM measurements is obtained by combining two local TM-to-SC displacement measurements, as those in LPF, with one SC-to-SC displacement measurement over a distance of 5 million kilometers[3]. The 10 pm/√Hz requirement is alleviated below 3 mHz, as, at low frequency, TMs motion due to parasitic forces is much larger than this measurement error.
- Identifying the key limits of the tracking performance, including a separation of noise from the phase measurement, which can be tested on ground, from the unwanted crosstalk from the large motion of the SC relative to the TM. This pickup is due to various metrological imperfections, which are shared with the LISA local interferometers, and can only be measured with a fully free-falling TM.

## 3 The in-flight experiments

Many details of the planned experiments have been given in [1] and [5], with a list of the main tests in [6]. We discuss here the status of the most important of these experiments, the measurement of the parasitic differential force noise, that accelerates the TMs out of their geodesics along the measurement axis x. Before doing that we need to summarize the basic features of the experiment.

Once in flight, LPF is a three-body (2 TMs and one SC) dynamical system, with all degrees of freedom (DoF) permanently controlled, except for the three translations of the system center of mass relative to the local inertial frame. With all displacements during measurement limited to less than one μm, we model this system as linear. With this we mean that we assume that its dynamics obeys, in the frequency domain, the following equations:

$$\vec{\mathcal{D}} \cdot \vec{q} = \vec{f} \quad \vec{f} = \vec{f}_d + \vec{f}_c \quad \vec{f}_c = -\vec{C} \cdot \vec{s} \quad \vec{s} = \vec{\mathcal{S}} \cdot \vec{q} + \vec{n} \tag{1}$$

The matrix $\vec{\mathcal{D}}$ describes the open-loop system dynamics, including inertial and elastic coupling terms. The vector $\vec{q}$ is formed by the measurable system generalized coordinates, and $\vec{f}$ is the generalized force/torque vector. $\vec{f}$ is the sum of the forces $\vec{f}_c$ due to controls, and of the remaining "direct" forces $\vec{f}_d$. $\vec{C}$ is the control matrix that act on the signal vector $\vec{s}$ to generate $\vec{f}_c$. Finally the signal matrix $\vec{\mathcal{S}}$ converts coordinates into signals, and $\vec{n}$ is the readout noise. The dependence of all quantities on frequency is omitted for simplicity.

The dynamics along the x-axis involves two DoF, $x_1$ and $x_{12}$. Ideally this dynamics is uncoupled from that of other DoF, the coordinates and signals of which we call $\vec{q}'$ and $\vec{s}'$ respectively. However misalignments, and other imperfections, introduce some coupling. It can be calculated that the effect of this coupling is to add, to the forces acting along x, an extra force term:

$$\vec{f}_d^{ct} = -\vec{\delta C} \cdot \vec{s}' - \vec{\delta \mathcal{D}} \cdot \vec{q}' \tag{2}$$

Eq. (2) holds up to linear terms in the "imperfection" matrix $\vec{\delta C}$, that expresses the pick-up, by the x-axis control loops, of signals $\vec{s}'$, and in the matrix $\vec{\delta \mathcal{D}}$, that represents the dynamical coupling of the motion along x to coordinates $\vec{q}'$. Within this linear approximation, both $\vec{s}'$ and $\vec{q}'$ are calculated in the absence of coupling. Thus in treating just the dynamics along x, a set of equations identical to eq. (1) holds, with $\vec{q}$ having just two components, $x_1$ and $x_{12}$, and provided that $\vec{f}_d$ also includes the cross-talk forces $\vec{f}_d^{ct}$ coming from the rest of the dynamics.

Once the TMs are set free, the control loops can never be interrupted, otherwise the system gets unstable. As a consequence, the forces $\vec{f}_d$ must be inferred from closed loop measurements of signals $\vec{s}$. By reshuffling eq. (1) we get:

$$\vec{D} \cdot \vec{s} = \vec{f}_d + \vec{f}_n \quad \vec{D} \equiv \vec{\mathcal{D}} \cdot \vec{\mathcal{S}}^{-1} + \vec{C} \quad \vec{f}_n \equiv \vec{\mathcal{D}} \cdot \vec{\mathcal{S}}^{-1} \cdot \vec{n} \tag{3}$$

The experimental goal will be to extract the differential, open-loop acceleration noise acting on the two TMs along the critical x axis, $(f_{x_2} - f_{x_1})/m$, one of the two components of $\vec{f}_d$. This can in principle be obtained from the data $\vec{s}$, by applying to them the matrix $\vec{D}$, representing the closed loop dynamics, provided that the elements of this matrix have been properly calibrated. Eqs. (3) also show that the measurement of $\vec{f}_d$ is corrupted by the unavoidable effect $\vec{f}_n$ of the readout noise $\vec{n}$.

For the matrix $\vec{D}$ we have a simplified linear model. This has been described in some details in [5]. The model contains a series of parameters: the absolute force calibration and the response times of actuators, i.e. of micro-thrusters and electrostatics, the static force gradient acting on each TM, the delays in actuation commanding, the interferometer cross-talk parameters. Some of these parameters are also assumed to be frequency dependent.

To derive all parameters, two different experiments will be performed on orbit, that consist of injecting a frequency-swept bias in turn into the DF loop and into the ES loop respectively, and to fit, for each experiment, the response template, expected from the model, to the output signals of both interferometers.

In [5] we have shown how, by combining the results from both experiments, with those coming from ground measurements of some interferometer parameters, the values of all model parameters can be measured. We also derived a Fisher-matrix based estimate of the accuracy of these measurements. The experiments have now been simulated with the end-to-end simulator of the mission [7]. This simulator was developed with the purpose of verifying the performance of the dynamical control system, and includes the complete non linear dynamics of the system, a noise model for disturbances, a full model for data transmission, etc.,. Details are given in the accompanying paper [8].

In Table 1 we report the results of the best fit procedure on the data from one simulation. Best fits are performed on whitened data, and can thus be tested for goodness by a standard $\chi^2$ test. The accuracy of the test is limited by the uncertainty on the knowledge of the noise PSD used to set up the whitening filter. Within this accuracy we don't find any significant discrepancy between our simplified model and the system response.

Comparing the parameter values obtained from the best fit, to those expected from the simulator is not straightforward. Indeed our model is simplified, compared to that of the simulator, and our parameter set cannot be fully mapped onto the much larger set used in the simulator. The comparison is possible for force gradients and for the absolute calibration of the electrostatic actuation forces, where the calculation from the simulator setting is more straightforward. For these parameters indeed the best fit matches with the expectations within the statistical errors.

We also notice that the precision of the parameters values estimated from the fit agrees with the analytical calculation reported in [5], a sign that the response of the system is linear to a good approximation.

Table 1. Parameters value from simulated experiments

| Parameter | Nominal value | Expected value from simulator settings | Estimated from data | Statistical Error |
|---|---|---|---|---|
| Absolute calibration of micro-thrusters. | 1 | $\approx 1$ | 1.0813 | 0.0005 |
| Absolute calibration of electrostatic actuation | 1 | 1 | 1.0000 | 0.0001 |
| Micro-thruster response time | 0 | < 1 s | 0.417 s | 0.002 s |
| Electrostatic actuation response time | 0 | < 1 s | 0.201 s | 0.003 s |
| Force gradient on TM1 | $-1.3 \times 10^{-6}$ s$^{-2}$ | $-(1.33 \pm 0.01) \times 10^{-6}$ s$^{-2}$ | $-1.319 \times 10^{-6}$ s$^{-2}$ | $0.004 \times 10^{-6}$ s$^{-2}$ |
| Force gradient on TM2 | $-2.0 \times 10^{-6}$ s$^{-2}$ | $-(2.04 \pm 0.01) \times 10^{-6}$ s$^{-2}$ | $-2.035 \times 10^{-6}$ s$^{-2}$ | $0.004 \times 10^{-6}$ s$^{-2}$ |
| Data bus delay within DF loop | 0 | < 1 s | 0.1997 s | 0.0003 s |
| Data bus delay within ES loop | 0 | < 1 s | 0.200 s | 0.009 s |
| Differential interferometer absolute calibration (measured on ground) | 1 | 1 | 1 | 0.0001 |
| DF reference interferometer absolute calibration (measured on ground) | 1 | 1 | 1 | 0.0001 |
| pick-up of SC motion by differential interferometer | 0 | $< 1 \times 10^{-4}$ | $1.2 \times 10^{-6}$ | $0.4 \times 10^{-6}$ |

Once the parameters have been obtained, displacement data can be converted into a force, and then analyzed to estimate the PSD of the force noise. We do this in the time domain by Fourier transforming $\vec{D}$ into its corresponding linear time-domain operator $\mathbf{D}$. In the time domain eq. (3) becomes then:

$$\mathbf{D}\left[\vec{s}(t')\right] = \vec{f}_d(t) + \vec{f}_n(t) \tag{4}$$

The main reason for performing time domain analysis is that the system dynamics is very slow. Response or relaxation times of up to tens of thousands seconds are commonplace. The common laboratory practice, of waiting the decay of all system transients before taking the necessary data, may take many hours or even days, and is not an option here. Thus force noise data must be extracted from displacement data, even in the presence of significant transients. Transients functions $\vec{s}_o(t)$ are solutions of the homogeneous equation associated to eq. (4):

$$\mathbf{D}\left[\vec{s}_o(t')\right] = 0 \tag{5}$$

Thus the operator $\mathbf{D}$ suppresses the transients, to within the accuracy with which the operator itself models the system dynamics.

An example of the results of this procedure on one series of simulated data is shown in Figure 1.

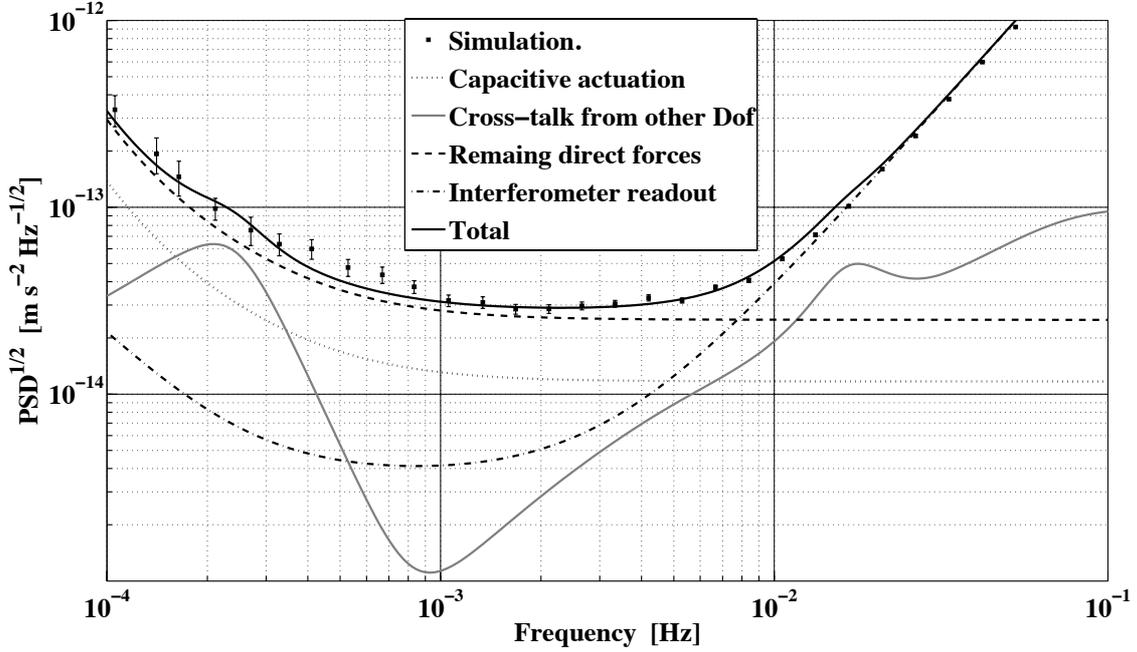

**Figure 1**. Square root of the PSD of differential force noise per unit mass on TMs, estimated from simulated displacement data as described in the text (filled squares). Lines represent the projected contributions to the force noise PSD based on the simulator inputs. *These have been chosen to coincide with worst case estimates for the various noise sources and significantly exceed the best current estimates, which are discussed in Section 5.* The agreement of the sum of these contributions (total) with the simulator data demonstrates the success of the simplified linear dynamics model of the LPF in quantitatively explaining the final experimental noise. See the legend for the meaning of the different curves.

The figure shows that the PSD of simulated data is in good agreement with that expected from our simplified linear model, if the same PSD is used for the different noise sources both in the model and in the simulator. However there are some minor, though statistically significant discrepancies at some frequencies. These are due to the already discussed differences between our model and the simulator, mostly with regards to the details of the control laws adopted for DF and ES along DoF different from x. We also performed a different kind of noise projection. We estimated the contribution of each noise source within the simulator, to the total noise in Figure 1. This was done by turning off all sources but the one under evaluation, and by estimating the total force PSD. We calculated then the root square sum of all these contributions, and the result was found in quantitative agreement with the total noise PSD. This confirms that, at least in the absence of signals, the system obeys the principle of superposition and thus behaves linearly.

Finally we note that, at this stage of the development, the noise part of the simulator is quite simplified. All force noise sources, except for those due to controllers or to actuators, are lumped into a single "direct forces" entry. In addition, *it is important to notice that worst case values are assumed for the PSD of the various contributions. These are in many cases much worse than their current best estimates*. These estimates are discussed the next section.

Finally we have to mention that the $\vec{f}_d$ does not exhaust the list of forces that are responsible for the differential acceleration noise of TMs. Part of the force is hidden, in eq. (4), in the non-diagonal terms of $\overrightarrow{\mathcal{D}}$. A differential force noise is indeed contributed by the product $\Delta\omega^2 x_1$, where $\Delta\omega^2$ is the difference between the static x-gradient of the force acting along x on TM2, and that acting on TM1. This term is implicitly calculated in the data reduction process discussed above, from the measured signal out of the $x_1$ interferometer, and the measured values of the gradient. Its PSD is found in good agreement with the expectations.

## 4  The performance budget

Given the demonstrated linear behavior of the system, the total force PSD can be calculated by giving the best estimate of the force PSD due to each independent source of disturbance, and then adding up the results. Many of these best estimates are now supported by experimental evidence obtained during the numerous test campaigns that have been performed. The best estimate of the mission performance budget, together with its topmost entries are shown in Table 2.

**Table 2 Leading sources of differential force-per-unit-mass disturbances, and their PSD values at 1 mHz**

| Source | PSD [fm s$^{-2}$/√Hz] | Estimated from |
|---|---|---|
| Actuation, x-Axis | 7.5 (0.8)* | Measurement of flight-model electronics stability |
| Brownian | 7.2 | Measurement with Torsion Pendulum |
| Magnetics | 2.8 | Measurement of magnetic field stability |
| Stray Voltages | 1.1 | Upper limit from torsion pendulum test campaign |
| Laser Radiation Pressure | 0.7 | Measurement of laser power stability |
| Force from dynamics of other DoF | 0.4 | From simulated dynamics of DoF other than x, and estimated worst-case values of $\overrightarrow{\delta\mathcal{D}}$ and $\overrightarrow{\delta\mathcal{C}}$ |
| Thermal Gradient Effects | 0.4 | Upper limit from torsion pendulum test campaign |
| Self-Gravity Noise | 0.3 | Upper limit from thermo-elastic stability simulations |
| Noisy Charge | 0.1 | Upper limit from charge simulation and measured voltage balance |
| Coupling to SC Motion via Force Gradients | 0.1 | From estimation of stiffness and simulated SC jitter |
| Total | 10.9 (7.9)* | Root square sum |

*The values within parenthesis refer to the free-flight mode. See text for explanation.*

Similarly, for the disturbances affecting the optical metrology the major contributions are listed in Table 3.

In Table 2 we also report the value of the actuation noise, and the resulting total acceleration noise, for a special experiment, the free flight mode [9], in which the ES along x is only intermittently applied. This experiment and its value to the LPF mission, will be further discussed later in the paper.

**Table 3 Leading sources of optical metrology disturbances, and their PSD values at 30 mHz**

| Source | PSD [pm/√Hz] | Remarks |
|---|---|---|
| Phase Noise | 4 | End-to end measurement on ground, including transmission through optical windows |
| Pick-up of motion along degrees of freedom different from x. | 1.6 | Analysis based on simulation of imperfections and measurement of alignments of optical bench |
| Total | 4.3 | Root square sum |

A large fraction of the entries in Table 2 and Table 3 are supported by experimental test. The ones that are not, in practical terms, can only be measured on orbit. This is the case for all the sources that involve the coupling to real SC-TM displacement, as for the force noise due to static gradients, for the spacecraft self-gravity, and for the metrology noise due to pick-up of motion of different degrees of freedom.

A summary description of the campaign of ground testing that supports the above budget, is described in the next section.

## 5   Estimation of LPF performance budget from ground testing.

As for the case of LPF, the ground testing also divides naturally into the campaign to assess the optical metrology performance, and that to estimate parasitic forces acting on the TMs.

### 5.1   *Ground testing of sources of stray force noise*

Most of the estimates for all sources of stray force noise in Table 2 are based on laboratory experience on ground, in almost all cases with prototype hardware representative of the final flight hardware. For some parameters, where the measurements on the ground are not possible, at least in practical terms, these are integrated by extended simulations based on the final flight configuration of the system. In the discussion that follows, we give the experimental and modeling evidence for the acceleration noise contribution at 1 mHz, as analyzed in detail for LISA Pathfinder. Where possible, we also give our best estimates for a given noise contribution at 0.1 mHz, which is relevant to the extrapolation of the LPF performance to LISA, which will be the subject of the next section.

Specifically:

#### *5.1.1   x- axis actuation*

Noisy actuation forces applied with the GRS electrodes opposite the TM x faces are the single most important LPF force noise source, dominated by the need to compensate the differential DC satellite self-gravity imbalance felt by the two TM. Any fluctuation in the amplitude of applied actuation voltages, generated by the GRS front-end electronics (FEE) produces force noise, increasing proportionally to the amplitude of the needed control forces[1]. LPF control scheme requires application of a force along x on TM2. In addition the same electrodes are used to control rotation around one of the axis normal to x (called z). This control is required for both TMs. Tests of the final flight electronics have measured relative actuation amplitude fluctuations at the 3-8 ppm/Hz$^{1/2}$ level at 1 mHz, largely uncorrelated between different electrode channels. The estimated 7.5 fm/s$^2$/Hz$^{1/2}$ differential acceleration noise considers the allotted gravitational balancing tolerances along x (0.65 nm/s$^2$) and the estimated gravitational torque around the z axis ($< 2$ nrad/s$^2$).

Among the various experiments planned on LPF, one, called the free-flight experiment [9], consists of letting the TM2 drift uncontrolled for intervals of several hundred seconds in between applied force impulses that put TM2 back into its initial state. This procedure is repeated many hundreds of times under closed-loop control, and data during the intervals of free-fall are analyzed to estimate the acceleration PSD in the absence of the actuation along x, a condition directly relevant to LISA. Analysis and simulations show [10] that the PSD is well estimated both below and above the pulse repetition frequency. In the absence of x actuation forces, the noise drops to 0.8 fm/s$^2$/Hz$^{1/2}$, due only to angular actuation. As the FEE instability is observed to increase approximately as 1/f in power, this figure raises to approximately 2.8 fm/s$^2$/Hz$^{1/2}$ at 0.1 mHz

*5.1.2  Brownian force noise*

Brownian force noise from residual gas damping has been discovered to be roughly an order of magnitude larger than previously estimated for the LISA TM[11]. The excess is largely due to the proximity – a 3-4 mm gap – of the TM to the surrounding GRS electrode housing, with dissipation created in the molecular flow in the narrow, high impedance channels around the TM [12][13]. The gas damping coefficient and resulting force noise have been estimated analytically and calculated accurately with numerical simulations for the LPF GRS geometry. Torsion pendulum measurements of pressure dependent gas damping have allowed quantitative verification of the model, at the 10% level, using LPF GRS prototype sensors in pendulum configurations sensitive to both forces and torques. The remaining uncertainty in the resulting LPF acceleration noise is tied to achieving the target residual gas pressure of $10^{-5}$ Pa inside the GRS vacuum chamber, which, however, has been demonstrated with representative prototype hardware and can be verified in the final pre-launch payload checkout. The result is the reported 7.2 fm/s$^2$/Hz$^{1/2}$. This contribution is frequency independent[12].

*5.1.3  Magnetics*

The magnetic force on each TM, $(\chi V/2\mu_o)\partial_x B^2/\partial x$, with V the volume, $\chi$ the susceptibility, and $\vec{B}$ the magnetic field, fluctuates because $\vec{B}$ fluctuates. Unfortunately, because of the quadratic nature of the effect, field fluctuations at all frequencies are important, as they are down-converted into the measurement bandwidth. In addition, while the susceptibilities of both TMs have been measured to be $< 2.5 \times 10^{-5}$ at DC, this increases to 1 around 600 Hz, where the effect saturates due to the skin effect. The mission prime contractor [14] has performed an extensive measurement campaign on magnetic field fluctuations, at frequencies from 0.5 mHz to 10 kHz, both from single components and from the entire spacecraft. The results have shown that high frequency fluctuations are barely detectable within the instrument noise, at levels well below 1 nT/√Hz. For the sake of noise budget estimation here, the instrument noise is taken as an upper limit.

At low frequency, the main electronic components were found to generate, at the TM location, magnetic fields on order of a few nT/√Hz at most. These values, with the proper margins, were used to estimate the PSD of 2.8 fm/s$^2$/Hz$^{1/2}$ reported in Table 2 for 1 mHz. The figure is contributed by the effect of interplanetary field that couples to the comparatively large static magnetic gradient caused by SC sources, by the fluctuation of local field and field gradient, and by the effect of the down conversion of high frequency field, that however, as stated, is just a measurement upper limit. For estimating the effect at 0.1 mHz, we note that the first contribution has been measured to have dependence on frequency as $1/f^{3/2}$, the second not faster than $1/f$, while the third can only be guessed, from the instrumental noise, to increase no faster than $1/f^2$. Assuming these frequency dependencies, the magnetic field noise increases to approximately $16 \text{fms}^{-2}/\sqrt{\text{Hz}}$ at 0.1 mHz.

*5.1.4  Stray voltages*

The dominant electrostatic disturbance for the LPF test mass is likely to be the interaction between the TM charge and the residual stray electrostatic field inside the GRS electrode housing. A fluctuating stray field will produce TM force noise by multiplying a non-zero average TM charge. A typical value for the TM charge is $10^7$ charges, which is the LPF discharge threshold and the expected charge accumulation in roughly one day. This noise source will give roughly 1 fm/s$^2$/Hz$^{1/2}$ differential acceleration noise assuming fluctuations in the average GRS stray voltage imbalance of 100 µV/Hz$^{1/2}$ at 1 mHz. Torsion pendulum measurements with gold coated metallic plates of similar dimensions to the LPF TM and GRS give upper limits of roughly 50 µV/Hz$^{1/2}$[15]. While the best published results measured inside a full LISA / LPF prototype GRS are of order 1 mV/Hz$^{1/2}$ [16], a current study to be published shortly has placed 100 µV/Hz$^{1/2}$ upper limits with a LPF prototype sensor at 1 mHz. Due to instrument noise, this upper limit increase by a factor 3 times at 0.1 mHz.

In addition to the surface effects above, the FEE actuation electronics will produce fluctuations in the electrode potentials, with a measured level of $10 \mu V/\sqrt{\text{Hz}}$, increasing at low frequency approximately as 1/f in power. This voltage adds to the intrinsic stray voltages discussed above.

From the frequency dependence of both contributions, we extrapolate a contribution from this source of $\approx 3.5 \text{fms}^{-2}/\sqrt{\text{Hz}}$ at 0.1 mHz.

*5.1.5 Laser radiation pressure*

Laser radiation pressure exerts a fluctuating force because of the amplitude instability of the laser. This has been measured to be at 50 ppm/√Hz at 1 mHz [17] at the actual power used of a few mW. The noise PSD is measured to increase as $1/f^2$ at low frequency.

*5.1.6 Force from dynamics of other DoF*

The dynamics of the other DoF may generate forces along the x axis on all three bodies, by two dominating effects. First electrostatic forces on both TMs, commanded by control loops that stabilize the other DoF, may have non-zero components along x, if $\overline{\delta C}$ eq. (2) has unwanted non-zero off-diagonal elements. The values of these have been estimated from the electrode geometrical tolerances and from the measured cross-talk between different channels of the FEE actuation electronics. To calculate the effects, one also needs to estimate the jitter of the forces commanded by the control loops. This has been obtained from the mission end-to-end simulator, and by the measured performance of the GRS and optical metrology displacement sensors.

The second effect is dynamical mixing $-\overline{\delta \mathcal{D}} \cdot \vec{q}'$, due to the coupling of the physical motion of the other DoF, via the off-diagonal terms of the dynamical matrix $\overline{\delta \mathcal{D}}$. These are in turn dominated by two main phenomena: non diagonal gravitation gradients and the rotation with the TM of static forces applied by control loops to compensate for the static gravitational forces. All gravitational fields have been calculated based on a detailed model of all the components of the SC, built on detailed measurements of the mass and locations of these components. This model is very fine grained (< 1mm) within the core assembly of the LTP, and becomes coarser (≈ cm) at SC level.

The calculated residual static gravitational field is then balanced by some proper balance mass. The error in this balancing procedure is taken as the maximum uncompensated force that the TM may experience in orbit and that the ES should then compensate for. Gradients are instead left uncompensated, and the estimated values are assumed in the calculation of $\overline{\delta \mathcal{D}}$. The final calculation of the dynamical cross talk, requires an estimate of the residual jitter of the coordinates of the other DoF. This is again obtained from the end-to-end simulator. Adding up both effects one gets the value of $\approx 0.4 \text{ fm s}^{-2}/\sqrt{\text{Hz}}$ reported in **Table 2**.

The frequency behavior of the PSD of cross-talk forces, depends on the details of the laws used for the control of the other DoF. In the present configuration, the dominant contribution at lower frequencies is due to the SC attitude control, which is driven by autonomous Star-Trackers (STR) aboard LPF. This controller applies electrostatic forces to TMs, and thus contributes force cross-talk. The present controllers have not been optimized to reduce the noise below 1 mHz. The frequency dependence of its closed loop gain produces the peak around 0.2 mHz visible in Figure 1, that decreases somewhat at 0.1 mHz. With the actual estimate of the cross-talk coefficients and the measured STR noise, the value at 0.1 mHz is $\approx 19 \text{ fms}^{-2}/\sqrt{\text{Hz}}$. However, with different control laws it is possible to move the peak to a decade lower in frequency while still maintaining good performance at 1 mHz. With such a control law, the effect grows slowly as ≈ 1/f in power to $\approx 1.3 \text{ fms}^{-2}/\sqrt{\text{Hz}}$ at 0.1 mHz. Such a control law modification is currently under discussion.

*5.1.7 Thermal gradient effects*

The conversion of GRS thermal gradients into forces via the well modeled radiometric and radiation pressure effects and via the less understood temperature-dependent outgassing effect has also been well characterized by torsion pendulum measurements [18][19]. Direct measurements of the force created by a temperature difference across the GRS electrode housing indicate approximately 100 pN/K at 295 K and $10^{-5}$ Pa, roughly half of which is attributed to outgassing and will likely be reduced further with the more vigorous bakeout envisioned for the final LPF GRS. The figure used to calculate the value reported in **Table 2**, is based on a true worst case estimate of the possible electrode housing temperature difference fluctuations of several $\mu K/Hz^{1/2}$ at 1 mHz. This corresponds to the absolute GRS temperature fluctuation level, not the relevant noise in the temperature difference. Thus thermal gradient acceleration noise will be a very minor contribution for LPF.

While temperature fluctuations will certainly increase at lower frequencies, detailed thermal modeling at 0.1 mHz is not yet available for LPF, given the concentration on the 1 mHz requirement, and its relatively easy satisfaction for this effect. However, LPF will be equipped with thermometry with a measured resolution of $10\ \mu K/Hz^{1/2}$ [20] that serve as a test bed for the GRS thermal behavior and an anchor point for payload thermal modeling. Coupled with a foreseen in-flight measurement of $dF/d\Delta T$, the effect of thermal gradients could be subtracted from the LPF data, leaving a residual noise of order of the measurement noise, corresponding in turn, to an acceleration noise of $\approx 1\,\text{fm s}^{-2}/\sqrt{Hz}$.

*5.1.8   Spacecraft self-gravity fluctuations*

The SC may be subject to thermoelastic distortion because of fluctuation of heat inputs and temperature during operation. The distortion modulates in turn the gravitational field generated by the SC and its various components. An extensive time resolved thermal simulation as been run [14], to estimate the PSD of the gravitational field fluctuations down to a frequency of less than a mHz. The figure reported in Table 2 is the result of this analysis. Based on the frequency dependence of the PSD at 1 mHz, observed in the simulation, we assume a $1/f^2$ increase at lower frequencies to $\approx 1\,\text{fm s}^{-2}/\sqrt{Hz}$ at 0.1 mHz.

*5.1.9   Random TM charging*

The same charge – stray field interaction relevant for the effect discussed in 5.1.4, produces force as the noise in the TM charge, which will have a $1/f^2$ PSD for Poissonian cosmic ray charging, will interact with any steady average DC electrostatic potential difference to produce $1/f$ force noise. The budgeted differential acceleration for the LPF test masses of $0.1\ \text{fm/s}^2/Hz^{1/2}$ at 1 mHz assumes an effective single elementary charge rate of 1000 single elementary charge events per second and a 10 mV stray DC bias imbalance. Calculations of the typical cosmic ray and solar charging give effective charge rates of order 300 /s [21], thus the value used here includes a substantial margin. As for the voltage imbalance, laboratory measurements using prototype GRS hardware typically show uncompensated DC biases of order 100 mV. However these experiments have also demonstrated the ability to measure and compensate this imbalance to better than 1 mV[16] [19][22], with 10 mV imbalance fully consistent with drifts and monthly readjustments.

*5.1.10   Coupling to spacecraft motion via force gradients*

Jitter in the spacecraft control along the x axis, estimated with the simulator to be at the $0.2\ \text{nm}/Hz^{1/2}$ level at 1 mHz, couples into the differential acceleration signal via any differential "stiffness" or elastic coupling of the two TM to the spacecraft. This stiffness is estimated to be dominated by the electrostatic force gradients due to the x-axis actuation, and by gravitational gradient, at the level of $1.3\ \mu N/m$. Other sources of stiffness originating in the GRS – TM interaction have been measured, with a LPF prototype sensor, to be roughly 5% of this level, dominated by the well-modeled electrostatic spring associated with the capacitive position readout[16]. This acceleration noise level is thus estimated to be insignificant for LPF, at the $0.1\ \text{fm/s}^2/Hz^{1/2}$ level, with backup possibilities to reduce it even further if necessary, by electrostatic "tuning" of the differential stiffness to zero and by subtraction of noise using the measured satellite control error signal[5]. The contribution is expected to be dominated by the sensor noise at low frequency, increasing then like $1/f$ in power to $0.3\ \text{fm/s}^2/Hz^{1/2}$ at 0.1 mHz.

*5.1.11   Unmodeled forces.*

In closing this discussion of force noise acting on the LPF test masses, it is worth considering any unmodeled noise sources, particularly those originating in the TM – GRS interaction, which has been considered as a potential source of force noise, given the short, mm-scale separations and importance of surface effects. Torsion pendulum measurements of the force noise acting on a LPF-like TM inside a prototype LPF GRS electrode housing integrated with a fully active LPF-prototype FEE, allows placing an upper limit of $100\ \text{fm/s}^2/Hz^{1/2}$ for a LPF differential acceleration noise from non-modeled surface forces at 1 mHz [23][24]. Though not fully representative of the space environment, for temperature or charging environments for instance – for which there are dedicated models and ground tests, as discussed above – such measurements rule out a wide class of disturbances at a level insuring that the GRS is close to the LPF performance goals.

5.2   *Optical metrology*

*5.2.1   Phase noise*

At the time of writing of this paper, the complete laser system, and all the electronics units of the optical metrology have been delivered. The optical bench has also been delivered, though the photodiodes needs to be replaced because of a failure. Nevertheless it has been possible to perform an end-to-end test of the entire chain, by using an engineering model of the optical bench, and by substituting the TMs with piezo-motor driven flat mirrors.

The details of a similar campaign using engineering models are reported in one of the accompanying papers in this same issue of the journal[25], and the details of the campaign using flight models will be reported in a forthcoming paper. In summary the chain included the flight models of

- Laser unit
- Acousto-optic modulator used for the heterodyne interferometers
- Laser control electronics
- Interferometer phase-meter
- Interferometer signal processing computer

The entire laser system was included inside a thermally stabilized vacuum chamber. The optical bench was included in a different, thermally stabilized vacuum chamber.

Data transmission bus and its harness were a faithful replica of their flight model and this was also the case for the on-board computer, that finally collects the data and transmit them to ground.

The results are shown in Figure 2.

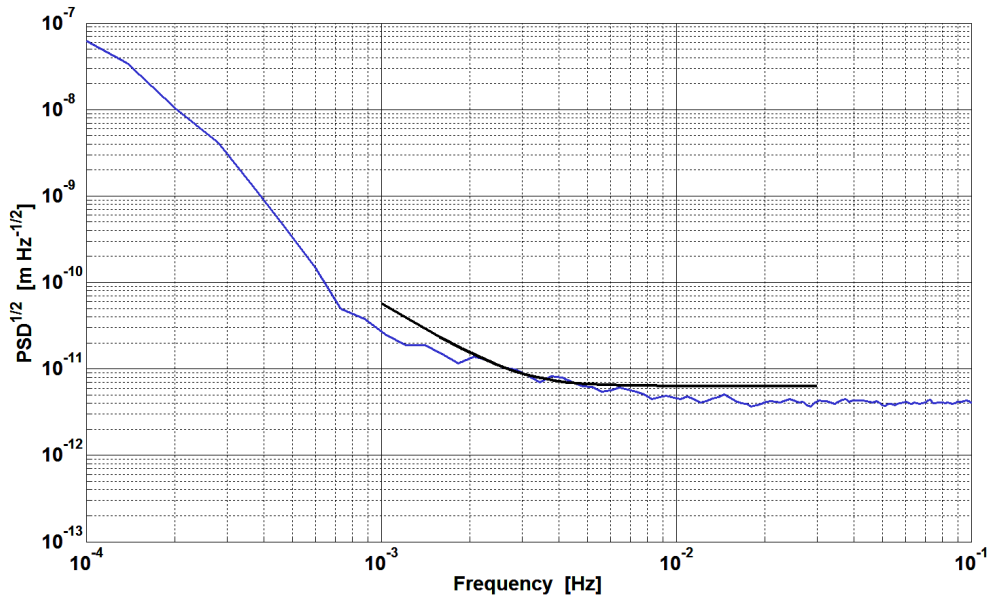

Figure 2 Square root of the PSD of the output of the $x_{12}$ interferometer within the flight model of the entire chain of the optical metrology. Only the optical bench is replaced by an engineering model. Blue line: experimental data. Black line: requirements. The increase below about 0.7 mHz is attributed to the laboratory environment.

The requirement for LPF are met within the specified frequency band with a white noise limit of $\approx 4\,\mathrm{pm}/\sqrt{\mathrm{Hz}}$. Below about 10 mHz the noise increases still remaining within the requirements. At the lowest frequency the noise increases rapidly. This is attributed to the laboratory environment. Measurements on the angular DoFs, which are appreciably more dynamically isolated from the environment, show a PSD increase as $\approx 1/f$ [26]. We predict that in the much quieter on orbit SC environment, also the measurements of linear displacements will perform similarly.

*5.2.2 Pick-up of motion along degrees of freedom different from x.*

As for the case of acceleration cross-talk, the assessment of this noise source by test requires the flight of LPF. However extensive measurements on the alignment of the optics have been performed that coupled with the above mentioned end-to-end simulation of the mission give a reasonable estimate of the expected PSD. This effect is expected to show-up mostly at higher frequencies.

# 6  Extrapolating to LISA

Two elements must be considered to fully exploit the applicability of the LPF results to LISA, which has a requirement for differential TM acceleration which is 7 times more stringent $\sqrt{2}\times 3\times 10^{-15}\,\mathrm{fm\,s^{-2}}/\sqrt{\mathrm{Hz}}$ at one decade lower in frequency (0.1 mHz). The first is the overall upper limit to stray force noise that LPF will provide, for any source of force noise, regardless of their origin or their inclusion in the noise budget. The second is the experimental assessment of the key parameters governing the dominant known noise sources, involving the space environment, the spacecraft hardware, and their interaction. In this section we give our current best estimate for how the extrapolation towards LISA will work and indicate where modifications to the overall LPF design will be needed to reach the LISA goals.

## 6.1  *Overall upper limit to non-modeled source of force noise*

The performance estimated in sects 4 and 5, if achieved on orbit, will put a firm upper limit on the acceleration noise for LISA at 1 mHz, roughly a factor 4, in power (2 in linear spectral density), above acceleration noise requirement for LISA. This would limit any unmodeled source of force noise to a level that would not threaten the ability of LISA to do unique and groundbreaking gravitational wave astronomy. Though LPF is only required to show performance above 1 mHz, the instrument noise – dominated by the optical metrology, with an $f^4$ conversion from phase noise power into equivalent acceleration noise power – is still sufficient to perform significant acceleration noise measurements at low frequencies.

Even assuming as a worst case the 0.1 mHz performance shown in Figure 2, the equivalent instrument limit from interferometry noise for differential TM acceleration is 30 fm/s$^2$/Hz$^{1/2}$. However, as discussed in sect. 5.2.1, in flight one expects a significantly better performance, better than the required $f^{-4}$ increase at lower frequencies. This limit would convert into a flat instrument limit for differential acceleration noise of 1.8 fm/s$^2$/Hz$^{1/2}$ (see dashed curve in Fig. 3).

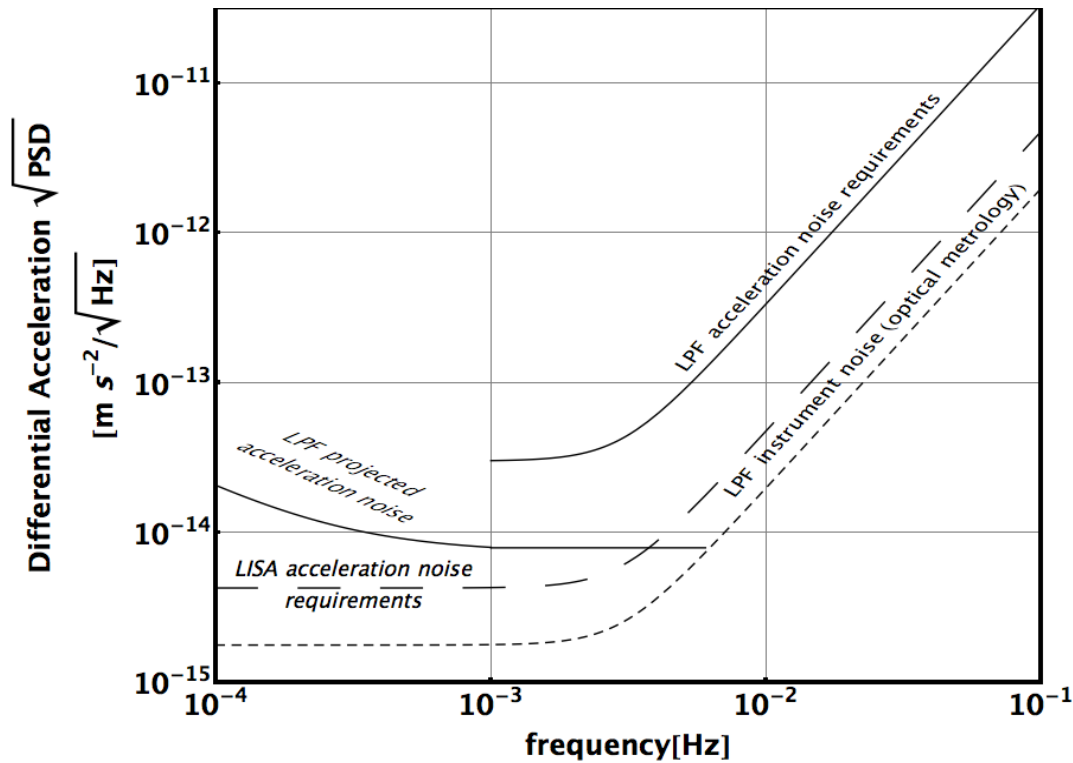

Figure 3 Projected differential acceleration noise performance of LPF in the free flight mode. The dashed line represents the optical metrology displacement noise converted into an equivalent acceleration noise. This line represents the limit of LPF ability to measure acceleration noise. Also reported are LISA and LISA Pathfinder acceleration noise requirements.

Simulations show the extra metrology noise introduced by crosstalk with the moving TM and satellite do not contribute appreciably to the noise at low frequencies.

From the extrapolation at low frequency of the various noise sources discussed in sect. 5.1, we expect known force noise sources to amount to roughly 23 fm/s$^2$/Hz$^{1/2}$ in the free-fall experiment at 0.1 mHz (see the solid LPF projected acceleration curve in Fig. 3) , a factor 5-6 above LISA requirements. With this background noise, any unmodeled noise source of this order should be visible.

## 6.2 *Modeled noise sources and requirements for LISA*

The general philosophy in noise budgeting known sources for LISA is that all differential acceleration noise sources should be kept below 1 fm/s$^2$/Hz$^{1/2}$ at 0.1 mHz, to keep the total sum below √2×3 fm/s$^2$/Hz$^{1/2}$ . Combining the LPF experiments with laboratory measurements on ground, we can summarize the status of known noise sources for LISA, what additional information is obtained from Pathfinder, and what improvements, in hardware or testing methods, are necessary for LISA.

### 6.2.1 *x- axis actuation*

In the absence of control forces along the x-axis, LISA should only have force noise from actuation of the torque around the z axis. LPF, with the current level of gravitational torque imbalance and FEE stability performance, is expected to reach $\approx 3 \,\mathrm{fm\,s^{-2}}/\sqrt{\mathrm{Hz}}$ around 0.1 mHz, which would use half of the total LISA noise budget. However in LPF, no attempt was made to improve the residual gravitational torque imbalance. Improvement by a factor of order 2-3 appears easily feasible. Additionally, we note that LPF will measure 6 rotational gravitational imbalances, giving a robust test of gravitational modeling and balancing capabilities. In addition to the improvement in gravitational balancing, it is certainly desirable that the FEE stability at the lowest frequencies improve by a similar factor, to make this contribution a minor entry of the budget. The current performance of the FEE is limited by various technical noise sources, and no fundamental limit was reached during its development. Thus reduction of the actuation noise to below 1 fm/s$^2$/Hz$^{1/2}$ appears achievable without any major redesign.

### 6.2.2 *Brownian force noise*

Reducing gas damping differential acceleration noise, from LPF's 7 to below 2 fm/s$^2$/Hz$^{1/2}$ across the band in LISA will require reducing the gas pressure from 10$^{-5}$ Pa to 5 10$^{-7}$ Pa or better. This can be reasonably achieved by replacing the LPF getter pump system with a tube that allows venting of the GRS vacuum chambers to space.

### 6.2.3 *Magnetics*

The projected magnetic field effect for LPF at 0.1 mHz is $16 \,\mathrm{fm\,s^{-2}}/\sqrt{\mathrm{Hz}}$ and needs to be reduced by a factor 10 for LISA. The effect is comprised of roughly equal contributions from several effects discussed in sect. 5.1.3. For the spacecraft in band and high frequency magnetic field noise, a significant overestimation of the noise is likely, due to instrument-limited testing and neglecting shielding factors for the AC components. Improved testing and, if necessary, dedicated shielding for magnetically noisy components, should allow a factor ten reduction for the spacecraft generated magnetic noise budget, both in-band and at higher frequencies. Additionally, the LPF static field gradients are dominated (at the 12 µT/m level) by contributions, now well identified, by unexpectedly magnetic thermal sensors on the GRS, with other sources an order of magnitude lower. This source can be removed for LISA, reducing the coupling to the interplanetary (and spacecraft) field fluctuations by an order of magnitude. Finally, there is margin for reducing the coupling further by improving upon the TM casting process, as the LPF TM susceptibility (-2.5 10$^{-5}$) is an order of magnitude larger than values quoted in literature for the same alloy. We also note that the in-band field fluctuations aboard LPF will be accurately monitored during flight by magnetometers, allowing verification of the field noise. Thus, despite relaxed magnetic requirements for LPF that would not be compatible with the LISA performance, the necessary factor 10 reduction in the magnetic force noise necessary for LISA is feasible and can be verified.

### 6.2.4 *Stray voltages*

Reducing the interaction of TM charge and fluctuating electrostatic fields to below 1 fm/s$^2$/Hz$^{1/2}$ , from the upper limit of 3 fm/s$^2$/Hz$^{1/2}$ for LPF at 0.1 mHz, will require a similar factor 3 improvement of the current laboratory upper limits inside the LPF GRS at 0.1 mHz near 300 µV/Hz$^{1/2}$. Efforts are under way to increase the measurement resolution closer to the 50 µV/Hz$^{1/2}$ level for measurement on the integrated sensor, which would then allow confirmation of results obtained for the simplified geometry in [15]. An improvement of the contribution from the FEE stability at 0.1 mHz, a factor two from the roughly 30 µV/Hz$^{1/2}$ for each electrode on LPF, is also needed for LISA. While this contributes to the torsion pendulum force tests, the contribution will also be isolated, on ground, by dedicated electronics testing. Finally, a measurement of force noise from stray voltage fluctuations can also be performed aboard LPF, taking advantage of its superior force resolution.

*6.2.5  Laser radiation pressure*

Reduction of the laser radiation pressure noise, currently estimated to contribute roughly 7 fm/s$^2$/Hz$^{1/2}$ at 0.1 mHz, by a factor 10, to below 1 fm/s$^2$/Hz$^{1/2}$, looks feasible by reductions both in the light power used and in its relative instability. The current performance of LPF optical metrology is not limited by light power, allowing a substantial reduction of the power hitting the TM without degrading the performance. In addition improved amplitude control at low frequencies can be developed and tested on ground.

*6.2.6  Force from dynamics of other DoF*

In LISA the spacecraft attitude is not controlled with STR. It is instead controlled to a much better accuracy, using the wave-front of the laser beams coming from distant SC. Thus the large contribution of the STR noise is largely suppressed. In addition the current estimate of this effect is based on an extrapolation of the micro-thruster noise, scaling like f$^4$, which is overcautious in the absence of any low frequency measurement. If instead LPF confirms the roughly < f$^2$ measurements of thrust noise, this contribution would basically be reduced by a factor 2.

*6.2.7  Thermal gradient effects*

LISA will require that the temperature differences across the GRS be smaller than 10 µK/Hz$^{1/2}$ to keep thermal gradient force noise below 1 fm/s$^2$/Hz$^{1/2}$, even with a small (25%) reduction in the coupling coefficient dF/dΔT, due to the drastically reduced role of the radiometric effect at the reduced LISA pressure. LPF will provide both a verification of thermal modeling at LISA frequencies and, if at all needed, a test of the effect's subtraction. Thus this effect looks well under control for LISA.

*6.2.8  Spacecraft self-gravity fluctuations*

The verification of the thermal model by LPF will also contribute to anchor the prediction of the thermoelastic distortion noise for LISA. This effect needs a factor 3 reduction from the $\approx 3 \mathrm{fm\,s}^{-2}/\sqrt{\mathrm{Hz}}$ predicted for LPF at 0.1 mHz. LISA would then require a 3 times better thermal stability than that currently projected for LPF, at least around the areas of the SC where the thermoelastic distortion is largest. This appears not to be a major technical challenge and has been studied in details during various LISA formulation studies.

*6.2.9  Random TM charging*

Compensation of the residual static potential imbalance at the 10 mV level will reduce this noise source below 1 fm/s2/Hz1/2 for LPF at 0.1 mHz even with a cautious allotment for the effective charge rate, and this is already sufficient for LISA. In addition to the in-flight procedures for measuring and compensating residual DC biases, LPF will also allow verification of the LPF and LISA TM charging model at low frequencies, with long term charge fluctuation measurements. This will allow to reduce the factor 3 margin we carry in our present estimates. In addition, as the charge time series will be measured, subtraction of this effect from the data appears to be feasible at the lowest frequency. This possibility will also be tested on LPF. However, pending the results of these measurement we think that the current estimate for LPF is also a cautious one for LISA.

*6.2.10  Coupling to spacecraft motion via force gradients*

This effect is already very small. In addition the residual force gradient on TM in LISA is expected to be reduced by a factor 2 relative to LPF.

The combination of the described improvements, no one being major or requiring a change of design, will in summary improve the acceleration noise in LISA by a factor ≈ 6 relative to LPF, bringing it well within LISA requirements. This assessment carries some layers of margin. Just to pick one example, pressure in interplanetary space and after more than one year of cruise, it's likely to decrease even beyond the $5\times10^{-7}$ Pa level discussed above. This margin will be used to make the implementation of the mission simpler.

## 7   Conclusions

In conclusion we have shown how the results of LPF, combined with ground testing, will allow extrapolation to the LISA parasitic acceleration performance, with reduced risk and reasonable confidence.

**Acronyms**

DF   Drag-Free controller

DoF  Degree of Freedom

ES   Electrostatic Suspension controller

FEE  Front-End Electronics

GRS  Gravity Reference Sensor

LPF  LISA Pathfinder

LTP  LISA Technology Package

PSD  Power Spectral Density

SC   Spacecraft

STR  Star-Tracker

TM   Test-Mass

TM1  reference Test-Mass for the drag-free control

TM2  reference Test-Mass for the electrostatic suspension